\documentclass[aps,prl,floatfix]{revtex4}
\usepackage{graphicx,eucal,bm}
\begin{document}
\title{How winding is the coast of Britain ? Conformal invariance
of rocky shorelines}
\author{G. Boffetta$^{1,3}$,  A. Celani$^{2,3}$, D. Dezzani$^1$ and A. Seminara$
^{2,4}$}
\affiliation{
$^1$ Dipartimento di Fisica Generale, Universit\`a degli Studi di Torino and INF
N, Sezione di Torino, V. Giuria 1, 10125 Torino, Italy
\\ $^2$ CNRS INLN, 1361 Route des Lucioles, 06560 Valbonne, France
\\ $^3$ ISAC Sezione di Torino, C. Fiume 4, 10133 Torino, Italy 
\\ $^4$ Dipartimento di Fisica and INFN, Universit\`a degli Studi di Genova,
V. Dodecaneso 33, 16146 Genova, Italy}

\begin{abstract}
We show that rocky shorelines with fractal dimension $4/3$
are conformally invariant
curves by measuring the statistics of their winding angles from
global high-resolution data.
Such coastlines are thus statistically equivalent to
the outer boundary of the random walk and of percolation clusters.
A simple model of coastal erosion gives an explanation for these results.
Conformal invariance
allows also to predict the highly intermittent spatial
distribution of the flux of
pollutant diffusing ashore.
\end{abstract}
\maketitle

\section{Introduction}
Forty years after Mandelbrot's seminal paper [\textit{Mandelbrot}, 1967]
where 
the concept of fractional dimension was introduced, there is a compelling 
evidence of the fractal nature of many geographical phenomena, including
the shaping of shorelines [\textit{Goodchild and Mark}, 1987].
Statistically self-similar curves are characterized by their 
fractal exponent $D$. If we select two points on the curve
and measure their distance $L$ along the curve
(e.g. by walking a divider of given width)
on average this will be proportional to their Euclidean distance $R$  
to the power $D$, i.e. $L \sim R^D$, 
where $D$ can take values between $1$ and $2$.
Such curves are widespread in nature,
and often enjoy a much richer symmetry than mere global scale invariance.
This is the case of conformally invariant curves, whose statistics
is covariant with respect to local scale transformations,
i.e. coordinate changes that preserve the relative angle between
two infinitesimal segments.
Conformal invariance is a pervasive feature of two-dimensional physics,
from string theory and quantum gravity
to the statistical mechanics of condensed matter 
and fluid turbulence
[\textit{Polyakov}, 1970; \textit{Belavin et al.}, 1984;
\textit{Schramm}, 2006; \textit{Bernard et al.}, 2006, 2007].
A remarkable  consequence of conformal invariance is the high degree 
of symmetry that often allows to make substantial analytical progress
[\textit{Cardy}, 2005; \textit{Bauer and Bernard}, 2006].
Among the many characteristic features that make conformally invariant curves 
peculiar within the 
class of self-similar ones,
the former are also distinguished by the special statistics of the winding
angle about a point belonging to the curve itself (see Fig.~1b). 
The probability distribution
of the winding angle is Gaussian, and therefore specified only by its mean
(that is zero, i.e. curves do not have a preferred
winding direction, or chirality), and its variance, that increases 
proportionally to the logarithm of the distance from the reference point
with a proportionality constant which depends on the fractal dimension.
This provides a simple and useful diagnostics for conformal invariance of 
curves extracted from experimental or numerical data.

Here we show that rocky shorelines display a Gaussian distribution 
of winding angles with a logarithmic dependence of the variance
as expected for conformally invariant curves 
with the correct numerical prefactor.
We use conformal invariance to predict the statistics of the
flux of pollutant diffusing over shorelines, which characterized
by a strongly intermittent spatial distribution
which can vary dramatically between locations just a few hundred meters apart
(see for instance Fig.~1 of \textit{Peterson et al.} [2003] about 
the Exxon-Valdez oil spill).


\section{Statistical analysis of rocky shoreline}

Since the famous paper of \textit{Mandelbrot} [1967],
the west coast of Britain has become
the paradigmatical example of fractal shoreline. In Figure~1 
we show a satellite image of a portion of the western coast of Scotland 
along with its digitized shoreline, that is a polygonal approximation
to the real coastline. It is also displayed in a double logarithmic 
plot the fraction of pairs of vertices of the polygon that lie within a ball
of diameter $R$: the slope of this curve is the correlation dimension 
[\textit{Grassberger and Procaccia}, 1983],
that is very close to $4/3$ in this case. 

The shoreline shown in Fig.~1 is one example of the curves
extracted from the high-resolution, self-consistent GSHHS database
[\textit{Wessel and Smith}, 1996].
The complete database covers the world shoreline which has been 
partitioned into $11503$ segments
of length $\approx 200$ {\it km} with a resolution of about $200$ {\it m}.
The computation of fractal dimension (as in Fig.~1) for
each segment, gives different values of $D$ that depend on the 
geomorphological processes at work in that particular geographical area. 
We observe a fractal dimension close to $1$ for sedimentary shores while
for rocky coasts it is about $1.3$ or larger. The overall most
probable value is found to be $D \simeq 1.2$.  Within this large sample,
we have selected the $1146$ shorelines which present a correlation
dimension close to $D=4/3$ (with a tolerance of $5 \%$).
The capacity dimension for such curves, 
computed by a box-counting algorithm, yields 
a value consistent with the correlation dimension, pointing to the 
conclusion that these are truly fractal curves and not multifractals 
[\textit{Grassberger and Procaccia}, 1983].

The statistics of winding angles for rocky coastlines is shown 
in Fig.~2.
The winding angle $\theta$ is defined as the angle between the line joining
two points separated by a length $L$ along the curve 
and the local tangent in the reference point,
measured counterclockwise in $rad$ (see Fig.~1b).
Because our curves do not have a preferred direction, the mean winding
angle $\langle \theta \rangle$ is very close to zero while the variance 
is found to grow with $L$ according to the logarithmic law
predicted for conformal invariant curves
[\textit{Duplantier and Saleur}, 1988; \textit{Duplantier and Binder}, 2002;
\textit{Wieland and Wilson}, 2003]

\begin{equation}
\langle \theta^2 \rangle = a + {2(D-1) \over D} \ln L
\label{eq:1}
\end{equation}
Here $a$ is a constant that depends on the details of the 
definitions and whose actual value is irrelevant.
The numerical evaluation of the coefficient in (\ref{eq:1}) gives
$D \simeq 1.33$, i.e. very close to the direct measure of $D$.
Figure~2b shows that the probability
density function (pdf) of $\theta$ is very close to a 
Gaussian distribution for different separations in the logarithmic range. 
Winding angle statistics have been computed using different reference points
located along the curve: we have found no detectable dependence on this choice.

Values of the fractional dimension other than $4/3$ (e.g. $D=1.2$ and
$D=1.5$) fail to give such an impressive agreement with the prediction 
for conformally invariant curves, in the sense that the prefactor differs
significantly from the value predicted by (\ref{eq:1}). 
An explanation for the peculiarity of this value of $D$  is
provided by a simple model, introduced by 
\textit{Sapoval et al.} [2004], of mechanical erosion of rocky coasts. 
Of course, the real properties of rocky 
coast morphology are the result of several mechanisms acting on various 
space-time scales [\textit{Carter and Woodroffe}, 1994] and are beyond
the scope of this simple modeling.
The basic ingredients of this model are two: {\it (i)} the mechanical 
resistance of rocks to erosive processes, essentially determined by 
their structure, composition and by the slow corrosion process due
to chemical agents, is assumed to have a typical scale of variation 
of the order of hundreds of meters and to be essentially uncorrelated 
on larger distances; {\it (ii)} rocks that are more exposed to the 
action of waves have a larger propensity to be fragmented by 
mechanical erosion: for instance, an isthmus will be eroded more 
rapidly than the shoreline within a gulf. 

This model can be implemented 
on a two-dimensional lattice where the sites
represent regions of land or sea of dimensions about a hundred meters. 
To every point on the land is assigned a number 
that measures the resistance of the rock to erosion. 
Then, if the resistance of a land site adjacent to the sea
falls below a given threshold, it
will be eroded, and thus transform into a sea site.
Subsequently, the resistance values for land sites along the shoreline
are updated depending on the local conformation of the coast
[\textit{Sapoval et al.}, 2004].
This procedure
is iterated  until no further updates are necessary and
a stationary artificial shoreline is obtained (see Fig.~3a).  
The similarities of this model with the well-known problem of percolation 
[\textit{Stauffer and Aharony}, 1991]
are evident, as already pointed out  in \textit{Sapoval et al.} [2004].
Indeed, in presence of rule {\it (i)} alone the islands generated by 
the algorithm would be statistically equivalent to percolation clusters
--- except for the inner ``lakes'' present in the latter case --- 
and thus display a fractal dimension $7/4$. However, rule {\it (ii)} 
prevents the formation of deep gulfs and peninsulae with narrow isthmi, 
therefore reducing the shoreline to the outer boundary of
percolation clusters that is known to have fractal dimension $4/3$ 
[\textit{Grossman and Aharony}, 1986; \textit{Saleur and Duplantier}, 1987].
Further refinements of the model, including damping of sea-waves and slow 
erosive processes do not modify the main features described above.
As a consequence of the statistical equivalence between the artificial shoreline
and the external frontier of percolation clusters, the former inherits
the known conformal invariance of the latter.
In Figure~3 we show the numerical results for the artificial 
shorelines generated by the model, which confirm the theoretical expectations.

\section{Intermittency of diffusing pollutants}

By virtue of the rich symmetry underlying conformal invariance, 
many interesting results can be obtained analytically. As a 
remarkable example we consider here the evaluation of the flux of pollutant
diffusing ashore from a source located in the sea.  Transport and mixing
of tracers is a complex issue of paramount importance
from microscopic to planetary scales [\textit{Ottino}, 1989]. 
At the simplest level of description dispersion is modeled as pure diffusion.
In the present case, this may be justified by estimates of the horizontal
eddy-diffusivity in the ocean that yield a ratio about $0.1$ to $1$
between mean currents and velocity fluctuations over scales of
a hundred kilometers [\textit{Marshall et al.}, 2006].

Pollutant concentration $c$ is therefore assumed to be given by the 
solution of the Laplace
equation $\Delta c=0$ with a pointwise source in the ocean and absorbing boundary 
conditions on the coastline ($c=0$). 
This problem can be solved with the aid of conformal transformations
by mapping the region of interest (i.e. a region of sea bounded by the 
shoreline) into an infinite strip, solving the Laplace
problem in the new domain (now a trivial task), and mapping 
the solution back to the initial region. 

The upshot of the conformally invariant nature of 
the shoreline is that techniques borrowed from theoretical physics
enable to compute analytically the pollutant flux distribution 
$\phi=\partial c / \partial n$ at the boundary 
[\textit{Duplantier}, 2000; \textit{Duplantier and Binder}, 2002;
\textit{Bettelheim et al.}, 2005].
The main result is that the probability of observing a flux $\phi$ 
of intensity $\phi_0 (R/R_0) ^{\alpha}$ --- where $\phi_0$
is the rate of emission by the source, 
$R$ the size of the region where the flux is computed, and
$R_0$ the distance of the source from the coast --- is proportional
to $(R/R_0)^{-f(\alpha)}$ with 
$f(\alpha)= \alpha + \frac{(2D-1)^2}{4(D-1)}
\left[ 1- \alpha^2/(2\alpha-1)\right]$,
for $R \ll R_0$. Small values of the flux 
correspond to large values of $\alpha$, whereas the largest ones 
take place for $\alpha \searrow 1/2$. This can be understood by 
means of the geometrical interpretation of the variable $\alpha$ 
[\textit{Duplantier}, 2000].
Indeed, let us recall that the flux inside a wedge of opening angle $\theta$
scales exactly as $R^{\pi/\theta}$. The result above can thus be interpreted
as if the shoreline was made of  a random collection of wedges 
of size $\sim R$ and opening angles $\theta$ with probability 
$\sim R^{-f(\pi/\theta)}$. Large $\alpha$ and small fluxes are equivalent 
to small $\theta$, i.e. deep fjords in the shoreline. On the opposite,
as $\alpha$ reaches the minimum value $1/2$, the flux attains its maximum
value $\sim \phi_0 (R/R_0)^{1/2}$ corresponding to $\theta=2\pi$, that
is a needle-like cape. The average flux $\langle \phi \rangle$ is exactly
$\phi_0$. By means of a variable change from $\alpha$ to $\phi$ 
it is possible to derive the exact probability density for the flux.
Besides the exact form, it is interesting to notice that for 
$\phi \ll \phi_0$ the probability of observing a value $\phi$ of the flux 
scales as a power law:

\begin{equation}
p(\phi) \sim \phi^{-2+\frac{(2D-1)^2}{8(D-1)}}
\label{eq:2}
\end{equation}

This power-law dependence is a reflection of the strongly intermittent 
character of flux fluctuations.
In Figure~4 we show 
the flux of pollutant emitted for a source located $40$ km offshore
the coastline of Fig~1, together with its probability density.
This closely follows the theoretical predictions for small fluxes
over a range of several decades.
Note that similar arguments hold for the longitudinal flux of pollutant
diffusing along the shoreline under reflecting boundary conditions. 

In conclusion, we have demonstrated that world coastlines with dimension
$D\simeq 4/3$ are conformally invariant curves by measuring their 
winding angle statistics. The distinguishing feature of such random curves
is their high degree of symmetry which enables to compute analytically 
many statistical properties. We have focused our attention on the flux
of pollutant
diffusing toward the shoreline, however many other interesting results
could be relevant to geophysical applications. For instance, 
an archipelago of conformally
invariant islands (loops) would display a power law distribution 
$A^{-1}$ of the number of islands of area larger than $A$ with a known
prefactor.
These would be also characterized by a ratio between the average area and  
the average squared radius equal to $\pi D/(2D-1)$.
All these properties, and many others, 
are also shared by self-avoiding walks (polygons), 
i.e. closed random walks 
that never hit themselves. These have been conjectured to be
 conformally invariant curves with dimension $4/3$ 
{\em via} the equivalence with
 stochastic Loewner evolution curves SLE$_{8/3}$ 
(see \textit{Lawler et al.} [2004] for a review).  
Remarkably enough, self-avoiding walks were 
introduced by Mandelbrot as well, when 
he conjectured the (now proven) equivalence between them and
the external frontier of two-dimensional Brownian motion.
Today, in view of our results, all these curves reveal 
their unexpected and intimate
connection with the brilliant intuition by Richardson and Mandelbrot
about the fractal nature of world coastlines.

%
%

\begin{acknowledgments}
This work has been supported by COFIN 2005 Project No. 2005027808.
\end{acknowledgments}

\begin{figure}
\end{figure}

\begin{figure}[h]
\centering
\includegraphics[scale=0.25]{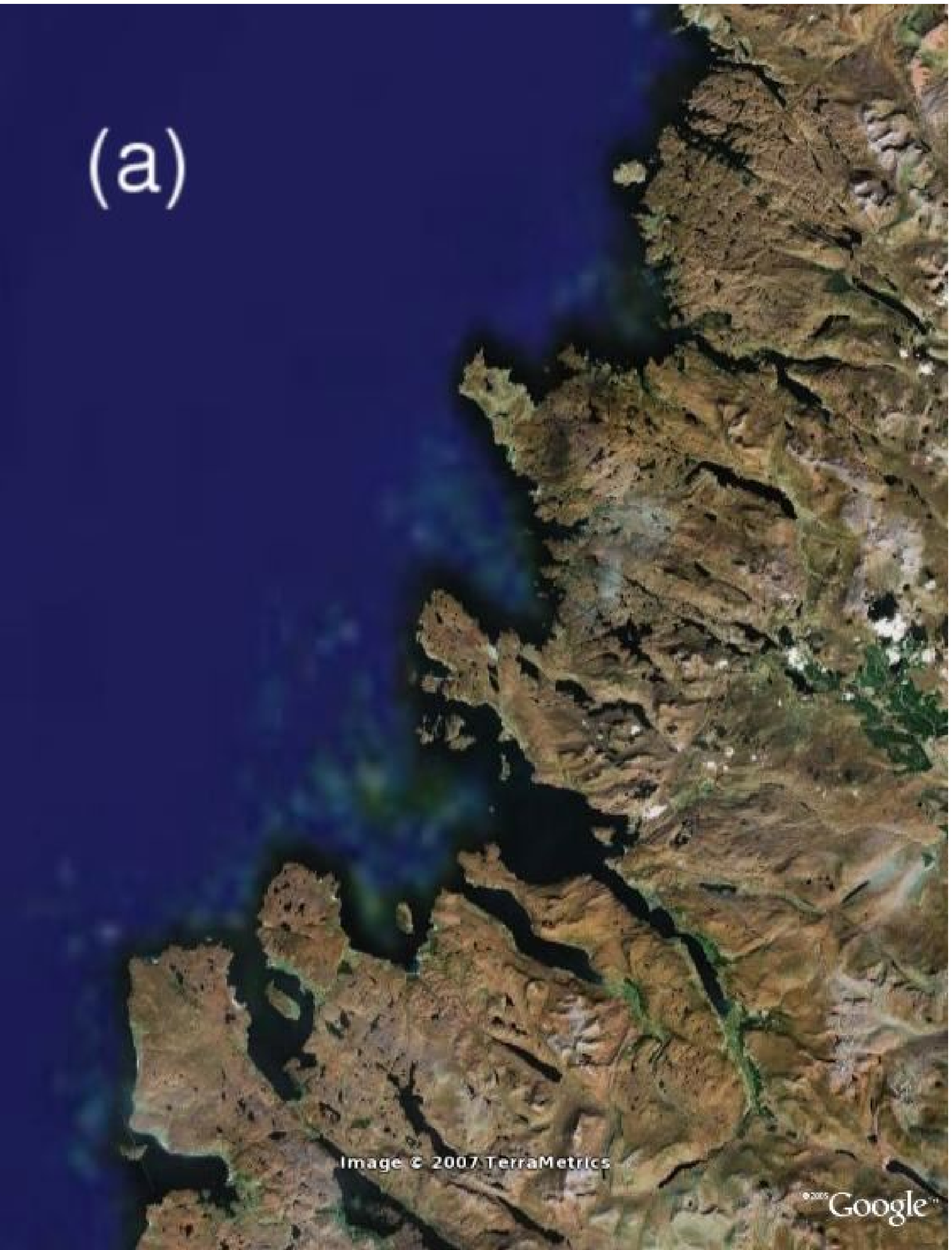} 
\includegraphics[scale=0.425]{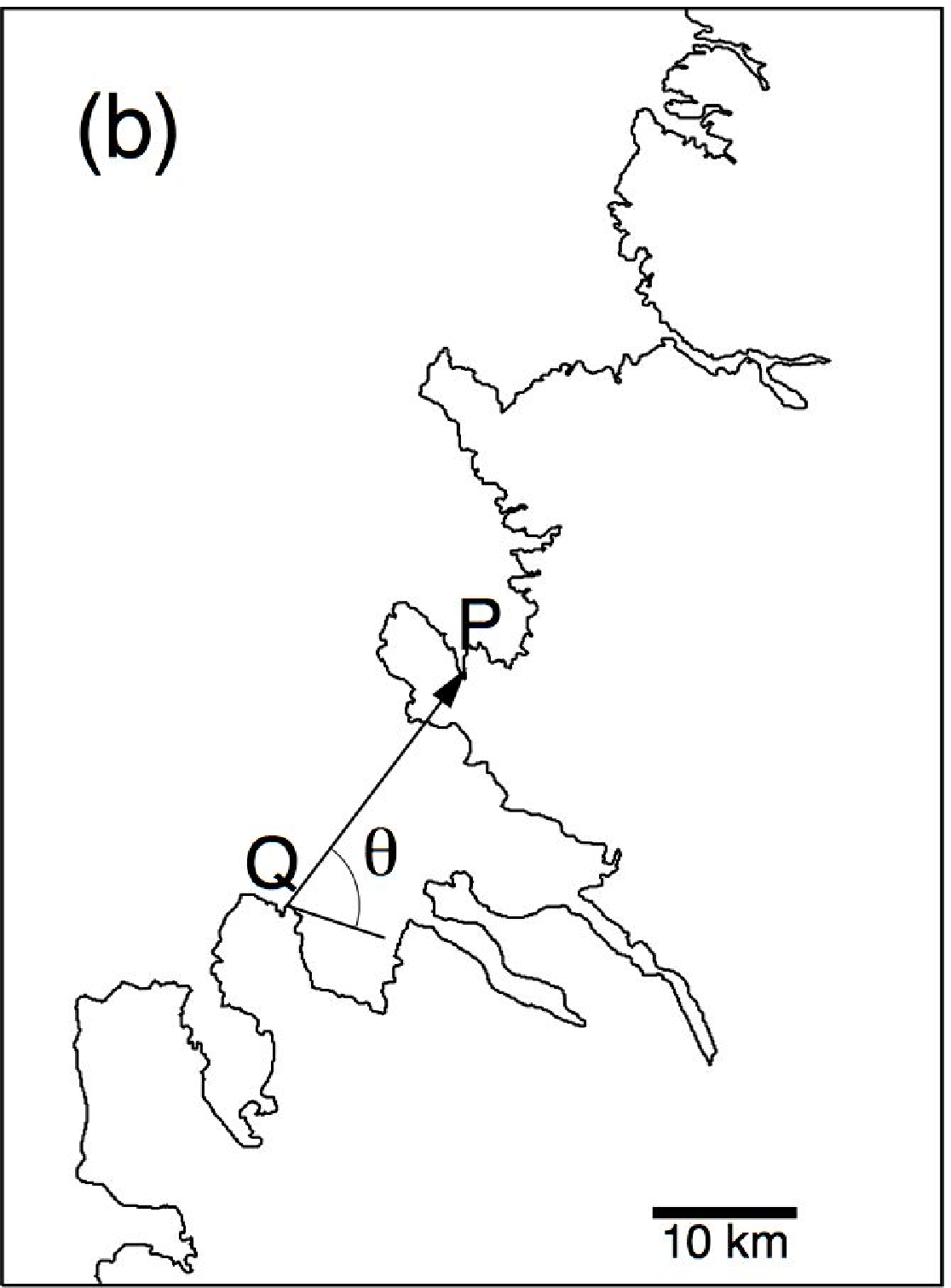} 
\raisebox{-1.0cm}{
\includegraphics[scale=0.44]{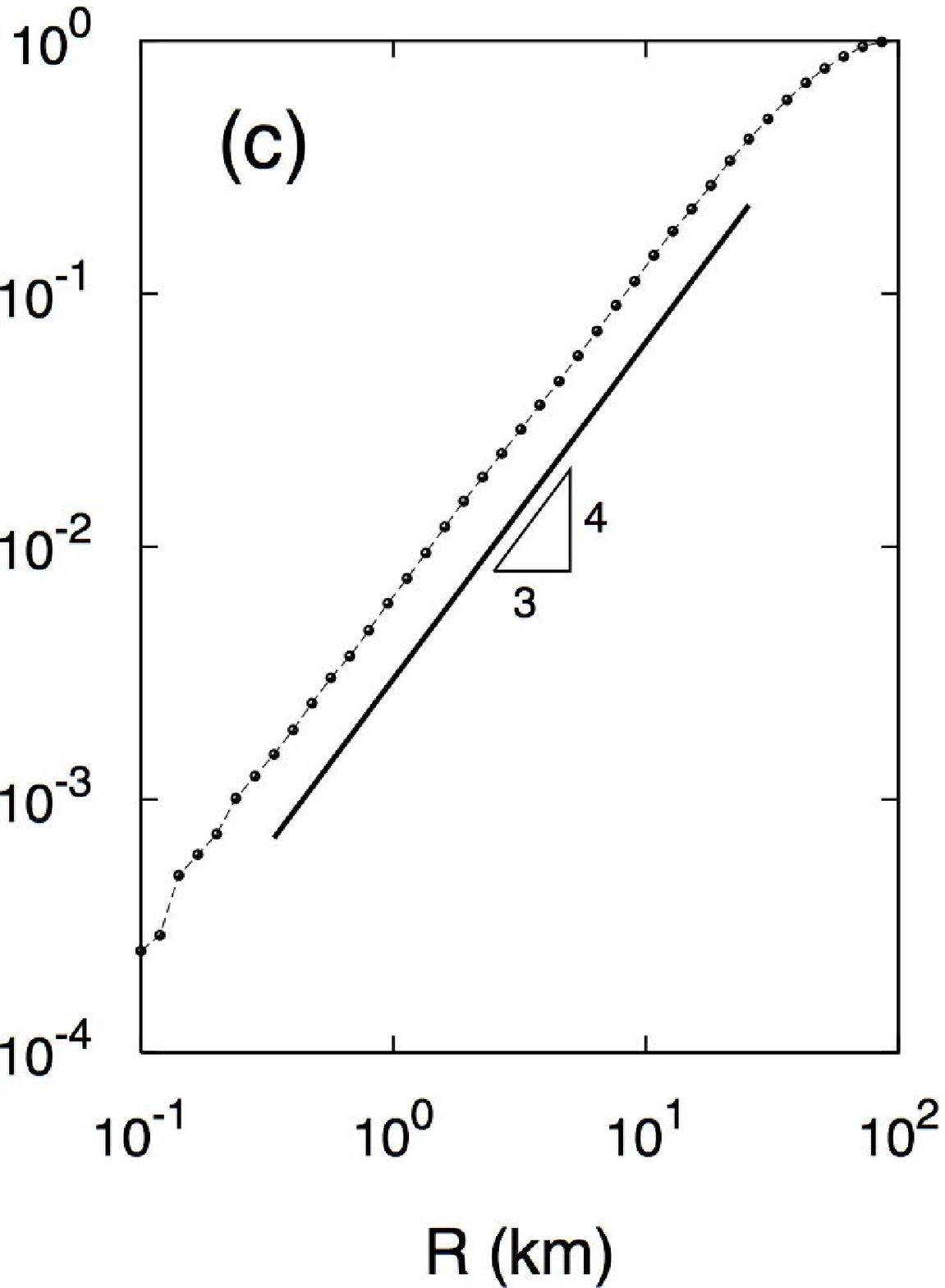}
}
\caption{The west coast of Scotland: an example of the $1146$ shorelines 
that have been searched for conformal invariance.
Panel (a) shows the satellite image of the geographical area, centered
around the point $58^{\circ}05'$N, $5^{\circ}21'$W. In Panel (b) is shown
the GSHHS polygonal approximation of the shoreline with resolution 
$\approx 200$ {\it m} together with an example of winding angle 
between points $Q$ and $P$.
Panel (c) shows the fraction of pairs of points of the curve (b)
lying at a distance smaller than 
$R$. The logarithmic slope of the curve is the fractal correlation dimension.
A least-squares fit for the data over the range from $300$ {\it m} 
to $20$ {\it km} gives an exponent $1.30 \pm 0.04$. 
Also shown for comparison a straight line of slope $4/3$. }
\label{fig1}
\end{figure}

\begin{figure}[h]
\centering
\includegraphics[scale=0.34]{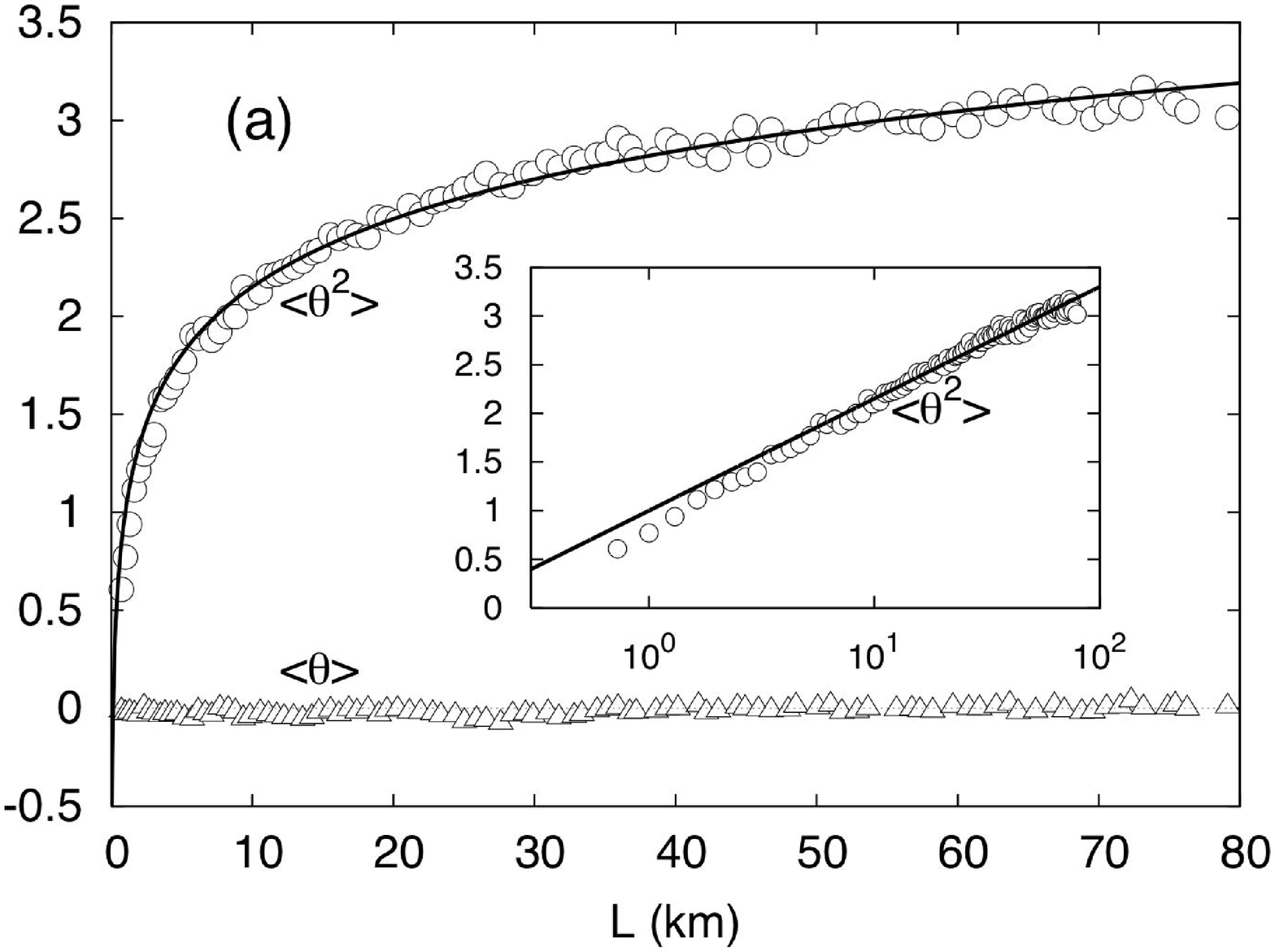}
\includegraphics[scale=0.34]{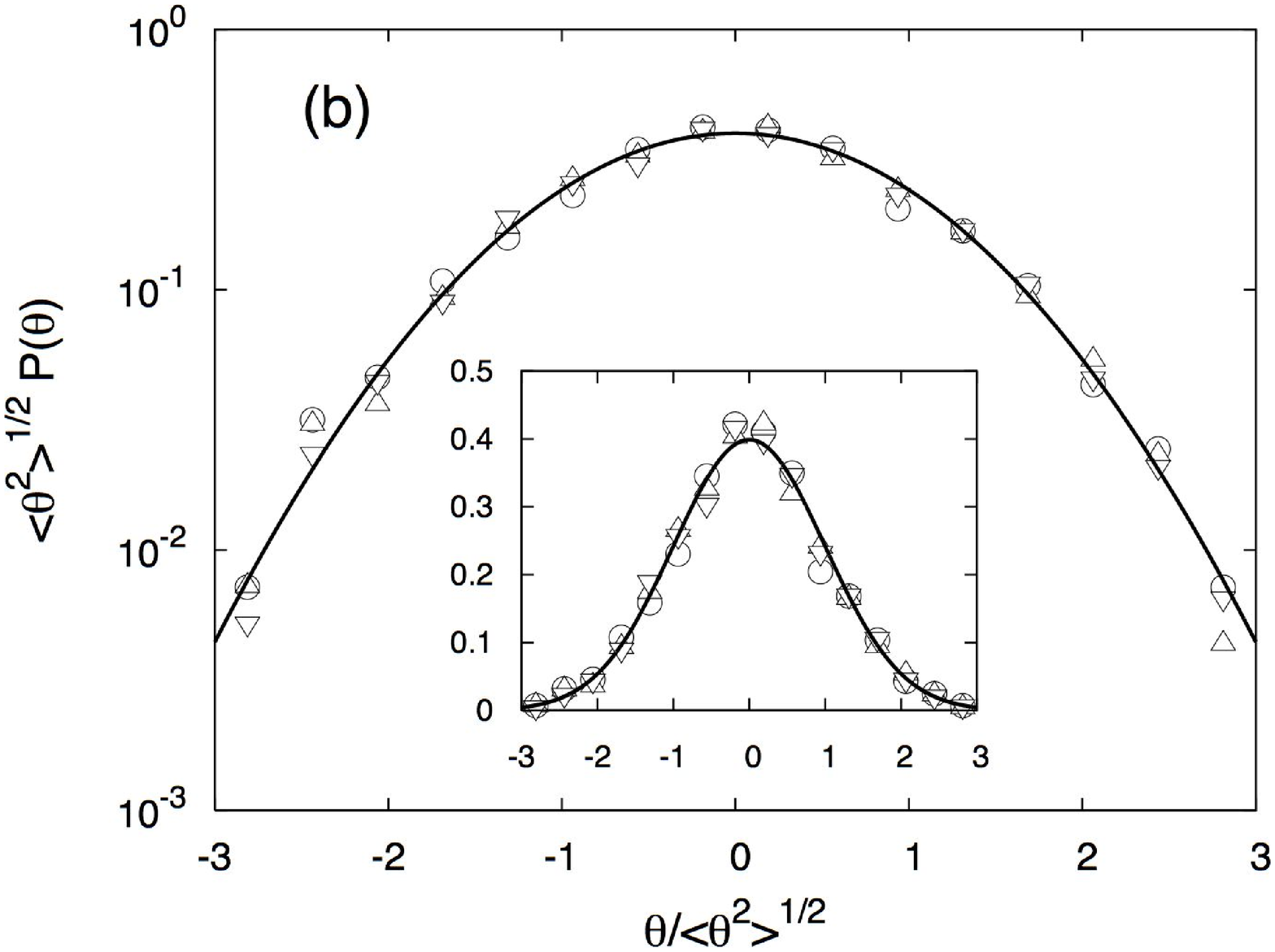}
\caption{Winding angle statistics. Panel (a) shows the mean and the 
variance of the winding angle as a function of the length of the shoreline 
between points $P$ and $Q$. 
(see Fig.~1).  The line is the law 
$\langle \theta^2 \rangle =a +\frac{2(D-1)}{D}\ln L$ 
with $D=4/3$ and $a=0.98$. In the inset,
the variance in semilogarithmic coordinates. In panel (b)  
is shown the probability density function
of the winding angle at lengths $L=5,10,20$ {\it km}
rescaled by the respective standard deviation and compared to the 
standard Gaussian density, in semilogarithmic (main frame) and in 
linear coordinates (inset).}
\label{fig2}
\end{figure}

\begin{figure}[h]
\centering
\raisebox{0.32cm}{\includegraphics[scale=0.50]{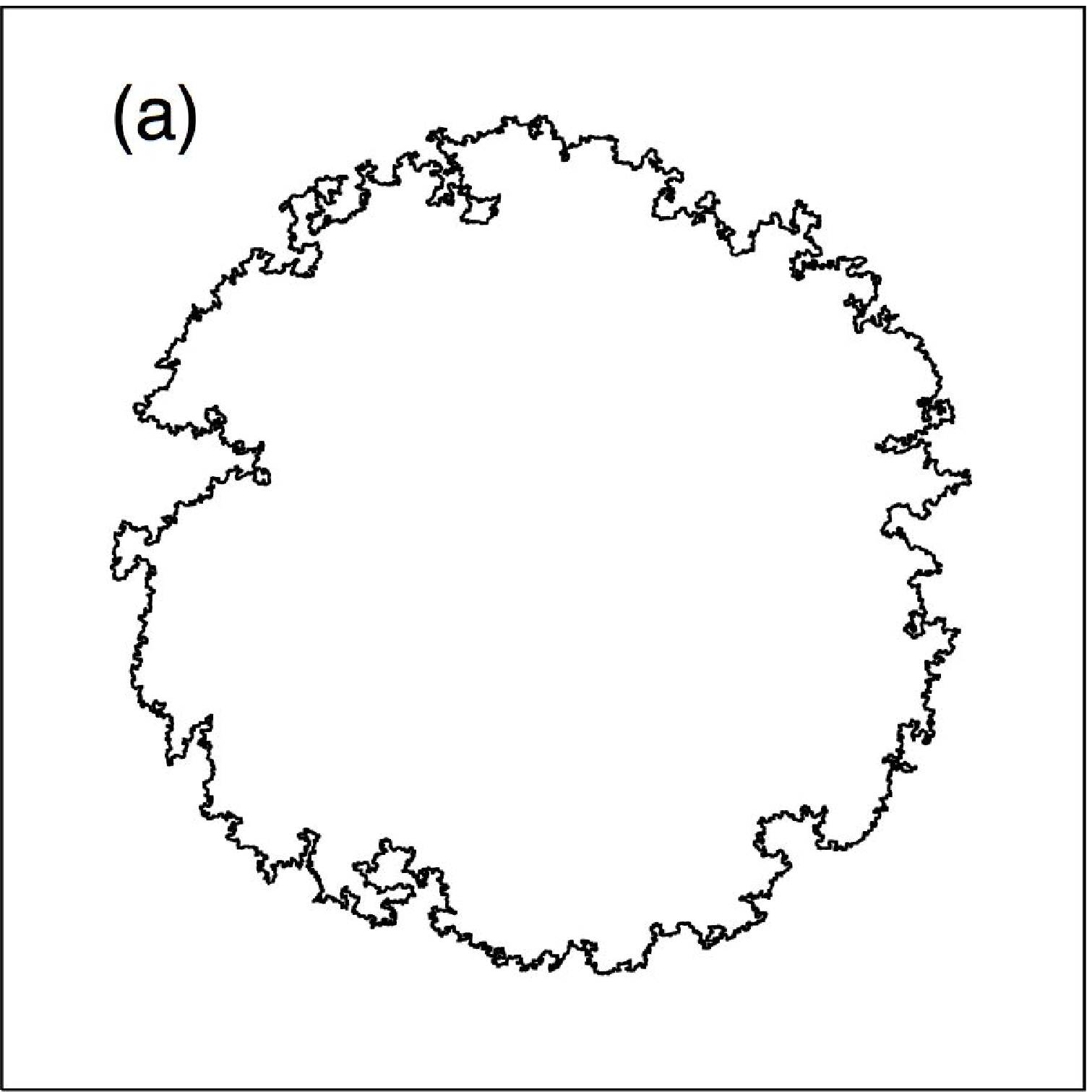}}
\includegraphics[scale=0.50]{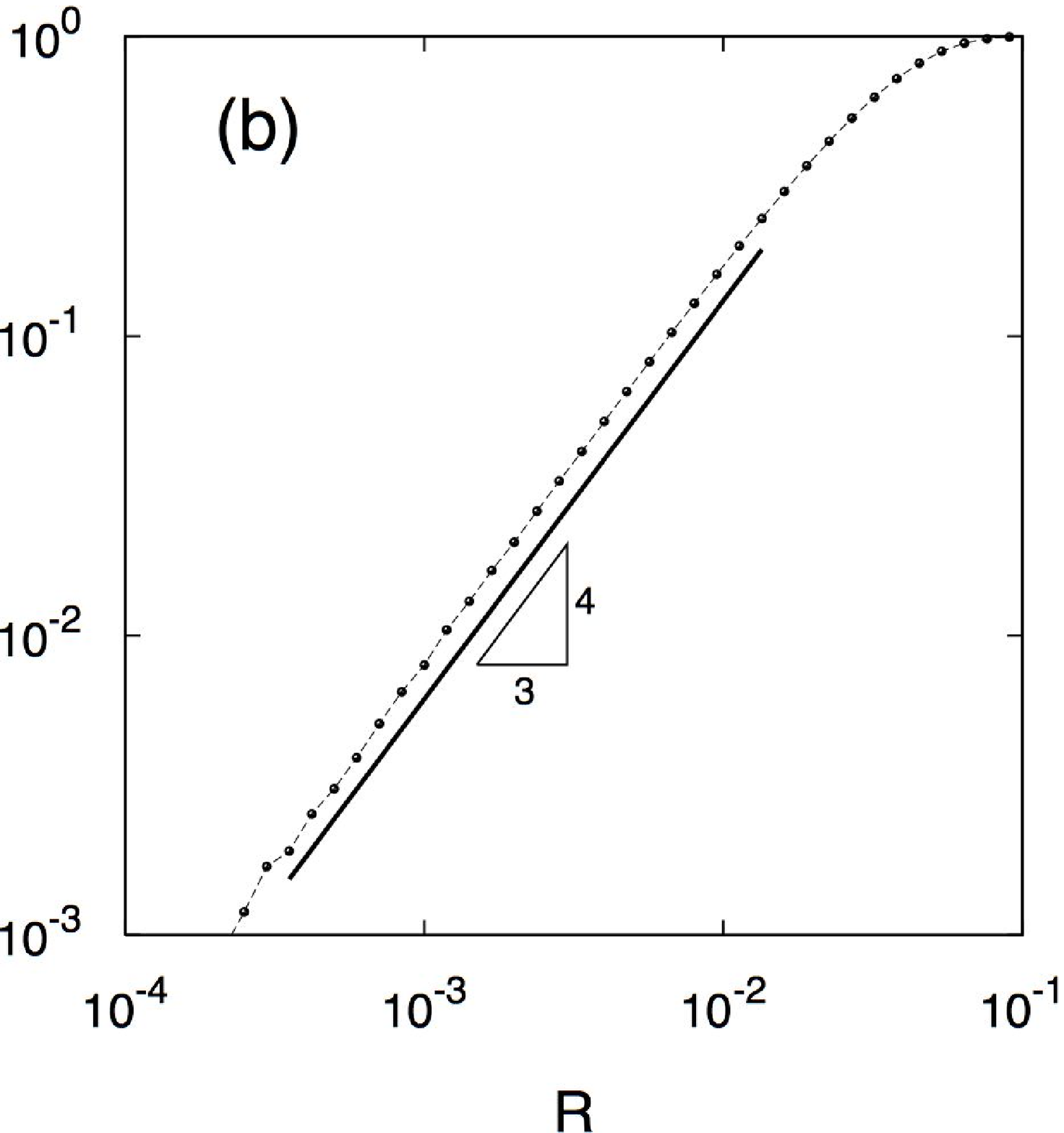}
\vspace*{0.2cm}\\
\includegraphics[scale=0.34]{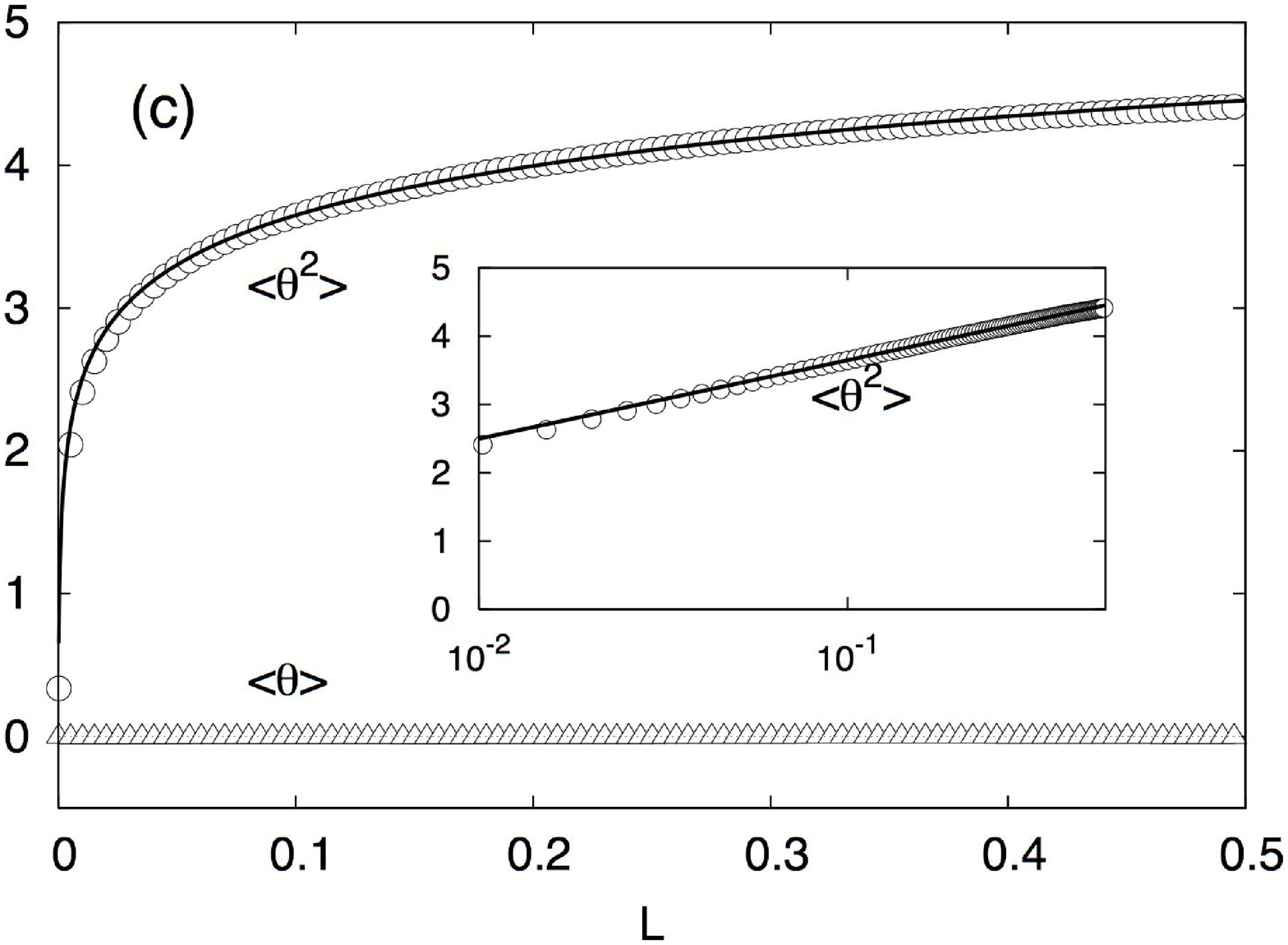}
\includegraphics[scale=0.34]{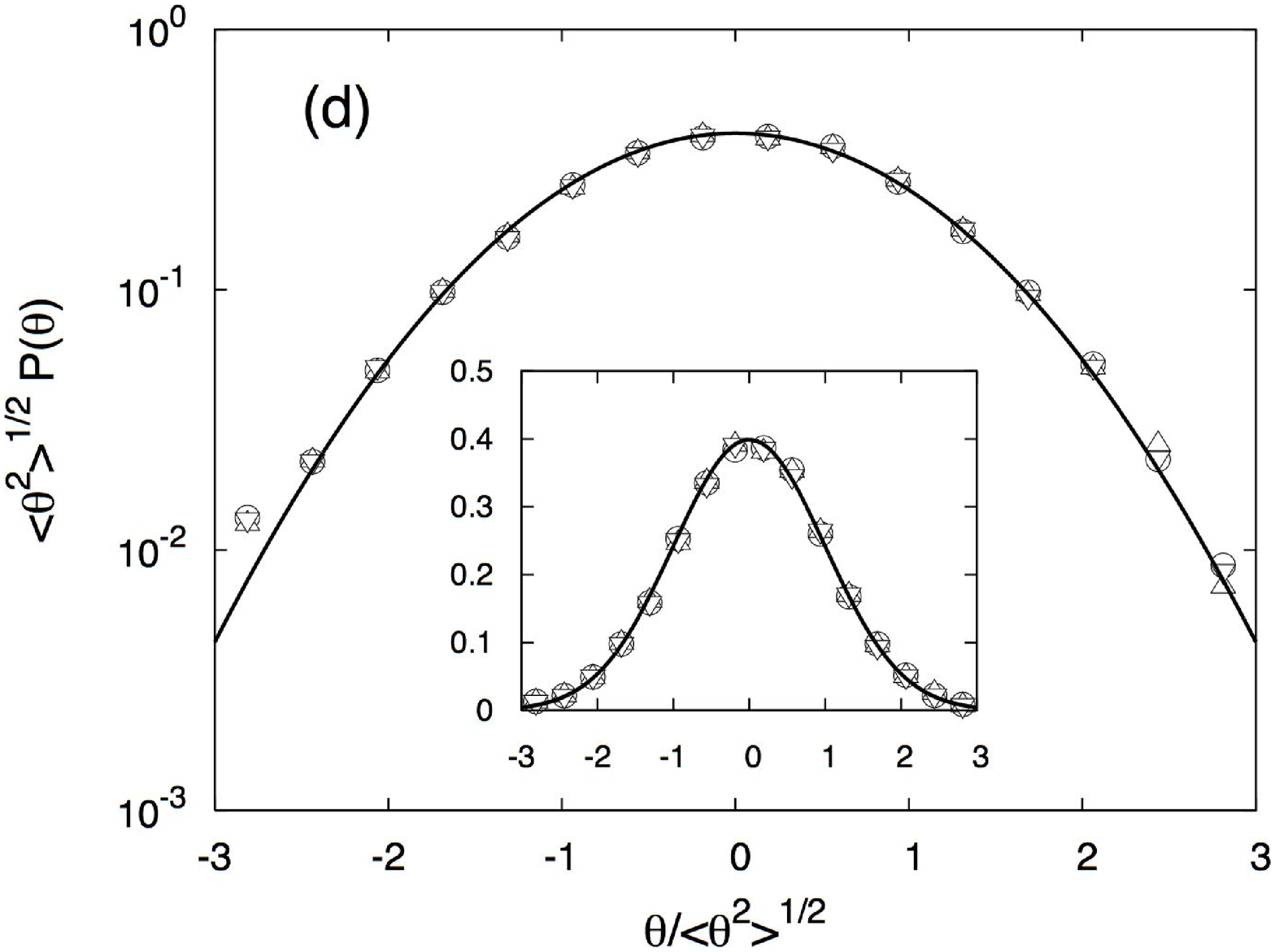}
\caption{Artificial shorelines. In panel (a) one example of a coast generated 
by the model of wave erosion described in the text. The simulation has been 
done on a square grid with $8000^2$ collocation points. 
The number of realizations is $800$. 
Panel (b) shows the correlation dimension. The unit for $R$ and $L$ 
is the simulation box size.
A fit in the range of $R$ between $5\cdot 10^{-4}$ and $0.1$ yields 
$D=1.32\pm 0.02$.
Panels (c) and (d) show the winding angle statistics as in Fig.~2
compared to the theoretical expectations.}
\label{fig3}
\end{figure}

\begin{figure}[h]
\centering
\includegraphics[scale=1.16]{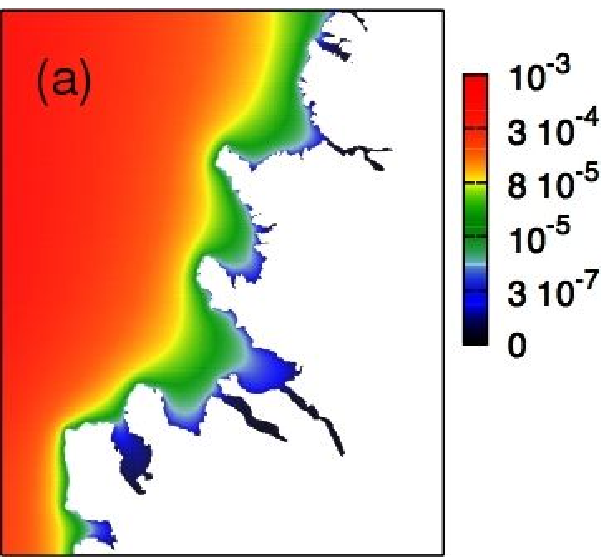} 
\includegraphics[scale=0.38]{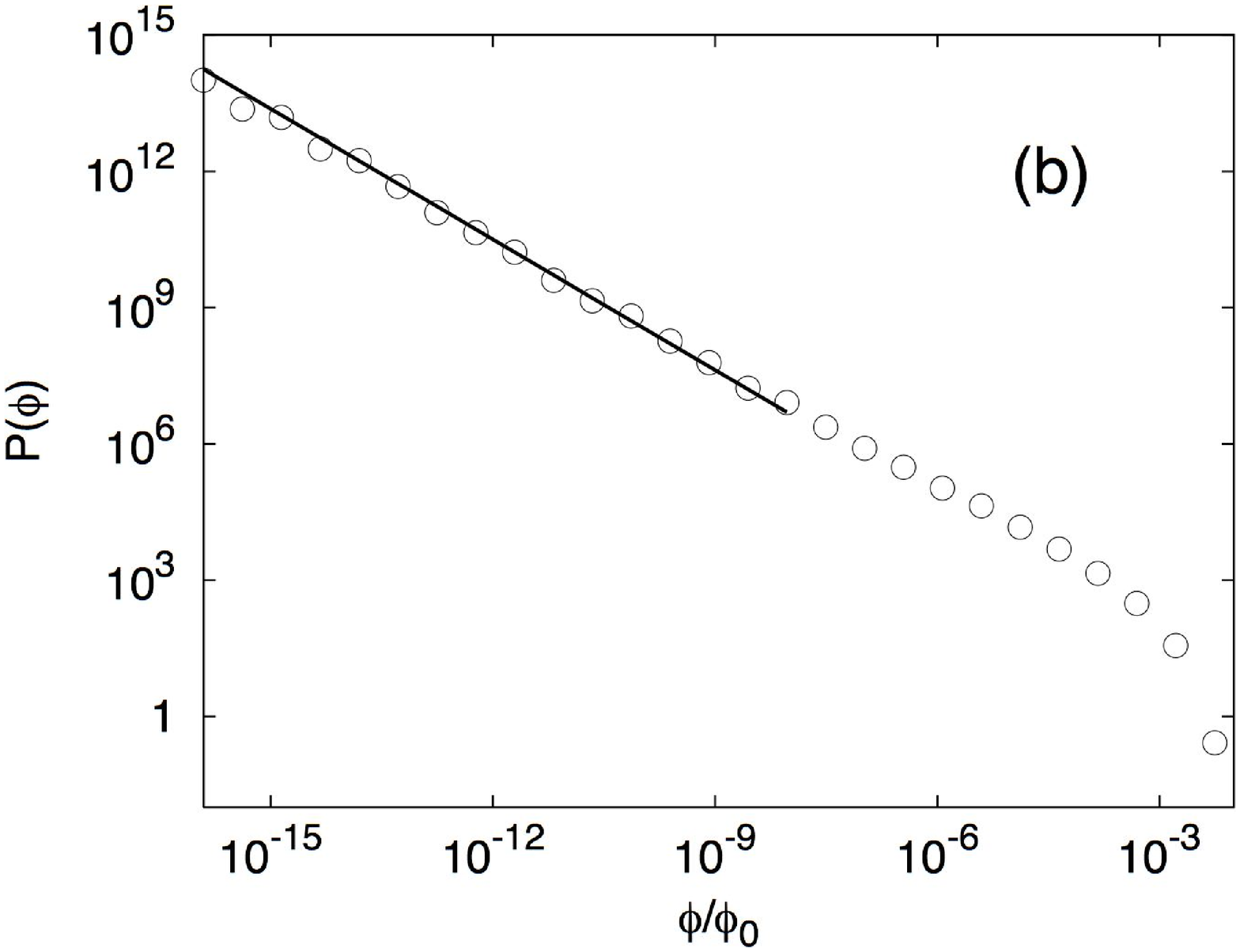}
\caption{Flux of diffusing pollutant. In panel (a) is shown the contour
plot of the pollutant concentration in the domain bounded by the shoreline of 
Fig.~1. Panel (b) shows the probability density of flux computed
along the present shoreline,
compared with the theoretical expectation (\ref{eq:2}) for the left tail.}
\label{fig4}
\end{figure}

\end{document}